\documentclass[fleqn,usenatbib]{mnras}

\usepackage{newtxtext,newtxmath}

\usepackage[T1]{fontenc}
\usepackage{ae,aecompl}
\usepackage{multirow}
\usepackage{subcaption}
\usepackage{enumerate}
\usepackage{xcolor, colortbl}
\definecolor{blueice}{rgb}{.85, .96, .94}

\usepackage{graphicx}	
\usepackage{amsmath}	

\title[Chemistry of sulfur chains]{Formation and desorption of sulfur chains (H$_2$S$_x$ and S$_x$) in cometary ice: effects of ice composition and temperature.}

\author[H. Carrascosa et al.]{
  H. Carrascosa,$^{1}$\thanks{E-mail: hcarrascosa@cab.inta-csic.es}
  G. M. Mu\~noz Caro,$^{1}$\thanks{E-mail: munozcg@cab.inta-csic.es}
  R. Mart\'in-Dom\'enech,$^{1}$
  S. Cazaux,$^{2,3}$
\newauthor
  Y. -J. Chen$^{4}$\thanks{E-mail: asperchen@phy.ncu.edu.tw}
  A. Fuente$^{1}$
\\
$^{1}$Centro de Astrobiolog\'{\i}a (CAB, CSIC-INTA), Ctra. de Ajalvir, km 4, Torrej\'on de Ardoz, 28850 Madrid, Spain\\
$^{2}$Faculty of Aerospace Engineering, Delft University of Technology, Delft, The Netherlands\\
$^{3}$Leiden Observatory, Leiden University, P.O. Box 9513, NL 2300 RA Leiden, The Netherlands\\
$^{4}$Department of Physics, National Central University, Jhongli City, Taoyuan County 32054, Taiwan\\
}

\date{Accepted 2024 July 17. Received 2024 July 17; in original form 2024 June 26}

\pubyear{2024}

\begin{document}
\label{firstpage}
\pagerange{\pageref{firstpage}--\pageref{lastpage}}
\maketitle

\begin{abstract}
The reservoir of sulfur accounting for sulfur depletion in the gas of dense clouds and circumstellar regions is still unclear. One possibility is the formation of sulfur chains, which would be difficult to detect by spectroscopic techniques. This work explores the formation of sulfur chains experimentally, both in pure H$_2$S ice samples and in H$_2$O:H$_2$S ice mixtures. An ultra-high vacuum chamber, ISAC, eqquipped with FTIR and QMS, was used for the experiments. Our results show that the formation of H$_2$S$_x$ species is efficient, not only in pure H$_2$S ice samples, but also in water-rich ice samples. Large sulfur chains are formed more efficiently at low temperatures ($\approx$10 K), while high temperatures ($\approx$50 K) favour the formation of short sulfur chains. Mass spectra of H$_2$S$_x$, x~=~2-6, species are presented for the first time. Their analysis suggests that H$_2$S$_x$ species are favoured in comparison with S$_x$ chains. Nevertheless, the detection of several S$_x^+$ fragments at high temperatures in H$_2$S:H$_2$O ice mixtures suggests the presence of S$_8$ in the irradiated ice samples, which could sublimate from 260~K. ROSINA instrument data from the cometary Rosetta mission detected mass-to-charge ratios 96 and 128. Comparing these detections with our experiments, we propose two alternatives: 1) H$_2$S$_4$ and H$_2$S$_5$ to be responsible of those S$_3^+$ and S$_4^+$ cations, respectively, or 2) S$_8$ species, sublimating and being fragmented in the mass spectrometer. If S$_8$ is the parent molecule, then S$_5^+$ and S$_6^+$ cations could be also detected in future missions by broadening the mass spectrometer range.

\end{abstract}


\section{Introduction}
\label{sect.introduction}
Sulfur chemistry in molecular clouds and circumstellar regions is a key factor to address the sulfur depletion problem. The cosmic abundance of sulfur is well-reproduced in diffuse clouds \citep{Jenkins2009_ApJ...700.1299J}. However, the abundance of sulfur atoms in the species detected in dense interstellar clouds is orders of magnitude lower than cosmic abundance \citep{Penzias1971_ApJ...168L..53P, oppenheimer1974_chemistry, bulut2021_gas_phase_abundances}. Therefore, a reservoir of sulfur atoms should be present in these environments. \cite{Caselli1994_ApJ...421..206C} hypothesized that S atoms could constitute a reservoir on the surface of dust grains. More recently, \cite{Laas2019_sulfur_depletion_model} used the depletion mechanism proposed by \cite{Ruffle1999_sulfur_depletion}, including the formation of S-bearing organic species to reproduce the abundances of sulfur in the interstellar medium. They concluded that sulfur may be relatively abundant in organic species, as the calculated abundances agreed with observations. Nevertheless, the molecular mechanisms leading to sulfur depletion remain unclear.\\

The formation of polysulfides (H$_2$S$_x$) is important in the astrochemical context, related to the sulfur depletion problem. Furthermore, these molecules can play an important role in biological systems, as they are present in microorganisms, acting as reducing agents in different metabolic routes. The formation and subsequent incorporation of polysulfides into biological systems is not known \citep{Kharma2019_review_polysulfides}. If present in comets, these molecules were delivered to the early earth, in particular during the Heavy Bombardment that ended 3.8 billion years ago.\\

Other studies by \cite{Jimenez-escobar2011AA...536A..91J} proposed that sulfur atoms may be present in allotropic forms of sulfur, being supported theoretically by \cite{Shingledecker2020ApJ...888...52S}. Hereafter, we will refer to H$_2$S$_x$ and S$_x$ molecules as polysulfides and sulfur allotropes, respectively. Furthermore, the more general term "sulfur chains" will be used to refer to both families of species indistinctly. Sulfur chains would be hard to detect by spectroscopic techniques. \cite{cazaux2022_H2S_chains} studied the formation of sulfur chains from pure H$_2$S ice samples submitted to radiation and warmup, both experimentally and using MonteCarlo simulations. They concluded that the formation of these species takes place in pure H$_2$S ice samples, and they may constitute a sulfur reservoir in the interstellar medium.\\

A dust grain travelling from the diffuse medium to a molecular cloud will experiment decreasing temperatures. Radiation and temperature in the diffuse cloud break any covalent bonds that may be formed, avoiding the formation of molecules. Between the diffuse and the dense cloud, known as the translucent phase, sulfur atoms can accrete on the surface of dust grains, what could contribute to the observed sulfur depletion \citep{cazaux2022_H2S_chains}. Once in the dense region, the lower radiation allow the formation and survival of molecules. Hydrogenation of oxygen and sulfur atoms impinguing the dust surface induce the formation of H$_2$O and H$_2$S molecules \citep{duley1980_formation_H2S_ISM}. Above H$_2$S thermal desorption temperature ($\approx$~90~K under laboratory conditions, or $\approx$~50~K in space \citep{Collings_2004_laboratory-to-astrophysical-context}) H$_2$S will not be retained in the ice mantle. Deeper in the cloud, temperature is lower, and both H$_2$O and H$_2$S will be adsorbed on the surface of dust grains, forming a mixed layer. However, if most of the H$_2$O has been incorporated to the ice mantle at higher temperatures, a H$_2$S-rich layer may be formed. Therefore, understanding sulfur chemistry in dense clouds requires the study of not only H$_2$S chemistry, but also H$_2$S chemistry in a H$_2$O matrix.\\

The Rosetta mission to comet 67P/Churyumov-Gerasimenko was devoted to understand the surface composition of comets, to shed light on their formation and evolution. The detection of sulfur species during the Rosetta mission by mass spectrometry was done using two techniques: 1) by an in-situ spectrometer, COSAC, on board of the Philae lander, and 2) using ROSINA, an orbiter spectrometer. COSAC detected molecules with a mass-to-charge ratio up to 60 \citep{Goesmann2015Sci...349b0689G}, therefore avoiding detection of any species containing more than one sulfur atom. Indeed, no S-species were identified in their data. However, the extended measuring range of ROSINA (up to 140 mass-to-charge units) and the multiple measurements of the desorbing species made it possible to measure the presence of molecules containing more than one sulfur atom \citep{Calmonet2016_detection_S3_S4_Rosetta, Mahjoub2023_sulfur_species_rosetta}.\\

\cite{Calmonet2016_detection_S3_S4_Rosetta} reported the presence of mass-to-charge ratios of 96 and 128, which were attributed to S$_3^+$ and S$_4^+$ cations. However, these authors could not confirm the presence of S$_3$ and S$_4$ molecules, as their detections could be originated by larger sulfur species being fragmented into S$_3^+$ and S$_4^+$ fragments. Mass spectra data for these type of molecules is required to better constrain the parent species resulting in these detections. Recently, \cite{Mahjoub2023_sulfur_species_rosetta} used ROSINA data to analyze the presence of sulfur-bearing species in the coma of 67P. They reported the presence of H$_2$S$_3$ molecules (i. e. mass-to-charge ratio of 
$\frac{m}{z} = 98$). Therefore, the combined data from \cite{Calmonet2016_detection_S3_S4_Rosetta} and \cite{Mahjoub2023_sulfur_species_rosetta} could imply the presence of H$_2$S$_3$ molecules which may be fragmented into S$_3^+$. Analogous processes could also be present involving larger H$_2$S$_x$ species.\\

\cite{cazaux2022_H2S_chains} and \cite{Altwegg_2022_sulfur_in_comets} made experiments on the formation of H$_2$S$_x$ and S$_x$ species containing up to 4 sulfur atoms. Apparently, there is no reason to constrain the chain length of sulfur chains in these experiments. However, the detection limits of the analytical techniques could not allow to determine whether large chains may be formed. Indeed, S$_8$, and other sulfur species in lower abundances, were reported in the residues made from similar experiments containing H$_2$S \citep{Tesis_guillermo}.\\

This work is focused in the understanding of the mechanisms leading to the formation and desorption of larger H$_2$S$_x$ and S$_x$ chains. The presence of these species may constitute a large sulfur reservoir which may help to decipher the sulfur depletion problem. Results are compared with works from \cite{Calmonet2016_detection_S3_S4_Rosetta} and \cite{Mahjoub2023_sulfur_species_rosetta}.\\

Thermal desorption temperatures for H$_2$S$_x$ species are determined in this paper, and their mass spectra are provided for the first time (Sect. \ref{sect.pure_H2S_ices}). Sect. \ref{sect.H2O_H2S_ices} studies the effect induced by a water-matrix surrounding H$_2$S molecules. In particular, the capability of sulfur species to react together forming sulfur chains in these astrophysical scenarios will be shown. The effect of temperature in the photoprocessing of the ice samples is discussed for both H$_2$S pure and H$_2$S:H$_2$O ice mixtures in Sect. \ref{sect.temperature_effect}. Sect. \ref{sect.comparison_rosetta} compares the experimental results obtained with reported data from Rosetta mission. Finally, conclusions and astrophysical implications are presented in Sect. \ref{sect.conclusions}.\\


\section{Experimental}
\label{sect.experimental_setup}

Experiments were carried out using the ISAC setup, at Centro de Astrobiolog\'ia (CAB, CSIC-INTA). ISAC is fully described in \cite{Guille2010}. Briefly, ISAC is an ultra-high vacuum chamber designed to mimic the conditions in the interstellar medium regarding pressure, temperature and UV radiation. The base pressure in ISAC is in the $10^{-11}$ mbar range, obtained by the use of turbomolecular and getter pumps. The lowest temperature, 10~K, which is measured on top of the sampleholder, is reached using a closed-cycle He cryostat, connected to a silicon diode. This system allows a precise control of temperature between 10~-~300 K, with an accuracy better than 0.1~K. Ultraviolet radiation is obtained using a microwave discharged hydrogen lamp (MDHL) from Opthos Instruments, and a controlled hydrogen flux with a pressure of 0.4~mbar. The MgF$_2$ window between the MDHL and the ice sample absorbs UV radiation for wavelengths shorter than 114~nm \citep{Asper2014}. For pure H$_2$S ice samples, H$_2$S (Nippon gases, 99,8~\%) was used, and highly distilled MilliQ water obtained from a Millipore water distribution system IQ-7000 was used for H$_2$S:H$_2$O ice mixtures.\\

Gas phase in ISAC is constantly monitored with a quadrupole mass spectrometry, using a Pfeiffer Prisma quadrupole equipped with a Channeltron detector. Molecules are ionized by electron impact at 70~eV. Solid phase is monitored after deposition and irradiation and during warm up of the ice samples with Fourier-transform infrared spectroscopy (FTIR), using a Bruker Vertex 70 spectrometer equipped with a deuterated triglycine sulfate detector (DTGS) in transmittance mode. For each measurement, 128 IR spectra were averaged with a resolution of 2~cm$^{-1}$.\\

Ice samples were deposited from the gas phase through a stainless steel tube oriented towards the MgF$_2$ substrate ($\approx$3~cm away). H$_2$S column density was fixed in the experiments, by growing the ice for 3~h at a constant pressure of 2$\times$10$^{-7}$~mbar, monitoring the ion current of $\frac{m}{z}~=~34$ with the QMS. Regarding H$_2$S:H$_2$O ice mixtures, H$_2$S was introduced at the same pressure, to ensure the same amount of H$_2$S molecules in the ice samples. H$_2$O was introduced through a different gas line, at the required pressure to obtain the 1:4 mixture desired, by monitoring $\frac{m}{z}~=~18$. The high UV absorption of H$_2$S molecules \citep{Gus2014}, together with the large column density fixed for this work determined the impossibility of irradiating after deposition of the ice, as only the top $\approx$~100 monolayers would absorb photon energy. Instead, simultaneous deposition and irradiation was used for the experiments. Temperature programmed desorption (TPD) experiments were also carried out at 1~K/min, devoted to detect the presence of so-formed photoproducts during their sublimation with the QMS.\\

Table \ref{Table.experiments} shows the different experiments carried out for this work. H$_2$O column density ($N$) was obtained from the IR spectrum after simultaneous deposition and irradiation of ice samples using a band strength ($A$) of $1.9 \times 10^{-16}$ cm molecule$^{-1}$ at 10~K and $2.1 \times 10^{-16}$ cm molecule$^{-1}$ at 50~K \citep{mastrapa2009_H2O}, and Eq. \ref{Eq.Band.strength}, where $\tau_{v}$ is the optical depth of the band, and $dv$ the wavenumber differencial. For H$_2$S, the available ice column density was obtained from the blank experiment using the IR spectrum and a band strength of $2.2 \times 10^{-17}$ cm molecule$^{-1}$ at 10~K \citep{cazaux2022_H2S_chains}, in agreement with the band strength provided by \cite{Yarnall2022_IR_band_strengths}. As the deposition time and H$_2$S partial pressure were kept constant for all the experiments, H$_2$S column density was assumed to be the same for all the experiments. This linear relationship between the column density of accreted ice and deposition time is presented in \cite{Cristobal2022_CO_density}, see Fig. 8 of this paper.\\

\begin{equation}
    N = \frac{1}{A} \int_{band}{\tau _{v}\; dv.}
    \label{Eq.Band.strength}
\end{equation}\\ 

To calculate the H$_2$S destruction in the experiments, H$_2$S column density after deposition and irradiation was obtained from the IR spectrum, using a band strength of $2.2~\times~10^{-17}$~cm$^{-2}$ at 10~K and $2.0~\times~10^{-17}$~cm$^{-2}$ at 50~K \citep{cazaux2022_H2S_chains}, subtracting it from the available column density calculated from the blank experiment, and considering the signal measured by the QMS for $\frac{m}{z} = 34$, shown in Table \ref{Table.experiments}.\\

\begin{table*} 
    \caption[]{Experiments performed in this work. $\frac{m}{z}~=~34$ column shows the normalized (with respect to the value in Exp. 1) integrated area recorded by the QMS during deposition of the ice samples for H$_2$S molecules. These values were used to calculate the total H$_2$S column density deposited during each experiment (\textit{initial column density}). $N$(H$_2$S) and $N$(H$_2$O) represent the column density measured after simultaneous deposition and irradiation of H$_2$S and H$_2$O, respectively. Last column shows H$_2$S destruction estimated from the calculation of the initial column density and the measured column density after photoprocessing of the ice samples.}
    \label{Table.experiments}
    \begin{tabular}{ccccccccc}
\hline
Exp.& Ice sample   & T  & Dep. time   &$\frac{m}{z}~=~34$    &Photon dose   &$N_f$(H$_2$S)  &$N_f$(H$_2$O) &H$_2$S destruction\\
             &              & (K)          & (min)             &(normalized)            & ($\times$10$^{17}$ cm$^{-2}$)  &($\times$10$^{17}$ cm$^{-2}$) &($\times$10$^{17}$cm$^{-2}$) &\%\\
\hline
\noalign{\smallskip}
\rowcolor{yellow!50}
\textbf{1}   &H$_2$S          &10 & 180 &1  & 0                       & 15    &0    &0\\
\noalign{\smallskip}
\textbf{2}&H$_2$S             &10 & 178 &1.2 & 6.4   & 7.3   &0    &58\\
\noalign{\smallskip}
\rowcolor{yellow!50}
\textbf{3}&H$_2$S             &50 & 182 &1.1 & 6.1   & 8.3   &0    &48\\
\noalign{\smallskip}
\textbf{4}&H$_2$S : H$_2$O (1:4)    &10 & 180 &0.9 & 3.9   & 6.1   &37   &53\\
\noalign{\smallskip}
\rowcolor{yellow!50}
\textbf{5}&H$_2$S : H$_2$O (1:4)    &50 & 186 &0.8 & 5.0   & 3.8   &46   &67\\
\noalign{\smallskip}
\hline
\end{tabular}\\
\scriptsize{}
\end{table*}

\section{Results and discussion}
\label{Results}

Generalising the synthetic pathway towards the formation of relatively short polysulfides reported by \cite{cazaux2022_H2S_chains}, the formation of H$_2$S$_x$ species could take place in subsequent additions of HS$\cdot$ radicals produced from UV irradiation to any H$_2$S$_x$ species:

\begin{eqnarray}
\label{Sch.HS}
    \rm{H_2S} \thinspace + \thinspace \rm{HS\cdot} \thinspace \xrightarrow{} \thinspace \rm{H_2S_2} \thinspace + \thinspace \rm{H\cdot}
\end{eqnarray}
\begin{center}
    ...
\end{center}
\begin{eqnarray}
\label{Sch.H2S3}
    \rm{H_2S_x} \thinspace + \thinspace \rm{HS\cdot} \thinspace \xrightarrow{} \thinspace \rm{H_2S_{x+1}} \thinspace + \thinspace \rm{H\cdot}
\end{eqnarray}

Hydrogen atoms produced from above reactions will diffuse in the ice, desorbing if they reach the surface, forming molecular H$_2$, or reforming H$_2$S molecules. However, it is also possible that radical-radical reactions take place:

\begin{eqnarray}
\label{Sch.H2S3}
    \rm{HS\cdot} \thinspace + \thinspace \rm{HS\cdot} \thinspace \xrightarrow{} \thinspace \rm{H_2S_{2}}
\end{eqnarray}

Not only HS$\cdot$ radicals are formed, but any other HS$_x\cdot$ radicals may be formed upon UV irradiation. Therefore, large H$_2$S$_x$ chains could be formed following (\ref{Sch.H2Sx}) or (\ref{Sch.H2Sx_radical_radical}):

\begin{eqnarray}
\label{Sch.H2Sx}
    \rm{H_2S_x} \thinspace + \thinspace \rm{HS_y\cdot} \thinspace \xrightarrow{} \thinspace \rm{H_2S_{{x+y}}} \thinspace + \thinspace \rm{H\cdot}\\
\label{Sch.H2Sx_radical_radical}
    \rm{HS_x\cdot} \thinspace + \thinspace \rm{HS_y\cdot} \thinspace \xrightarrow{} \thinspace \rm{H_2S_{x+y}}
\end{eqnarray}

Similar processes can be responsible of the formation of sulfur allotropes from sulfur atoms:

\begin{eqnarray}
\label{Sch.S}
    \rm{S\cdot} \thinspace + \thinspace \rm{S\cdot} \thinspace \xrightarrow{} \thinspace \rm{S_2} \thinspace
\end{eqnarray}
\begin{center}
    ...
\end{center}
\begin{eqnarray}
\label{Sch.Sx}
    \rm{S_x\cdot} \thinspace + \thinspace \rm{S_y\cdot} \thinspace \xrightarrow{} \thinspace \rm{S_{x+y}} \thinspace
\end{eqnarray}

Sulfur allotropes can be also produced from dehydrogenation of sulfur chains, what may be specially favoured if a cyclic allotrope (number of sulfur atoms greater 5) is obtained:

\begin{eqnarray}
\label{Sch.S_cyclic}
    \rm{H_2S_x} \thinspace \xrightarrow{} \thinspace \rm{H_2} \thinspace + \thinspace \rm{S_x}
\end{eqnarray}

\subsection{Pure H$_2$S ice samples}
\label{sect.pure_H2S_ices}

\begin{figure*}
  \centering 
  \includegraphics[width=\textwidth]{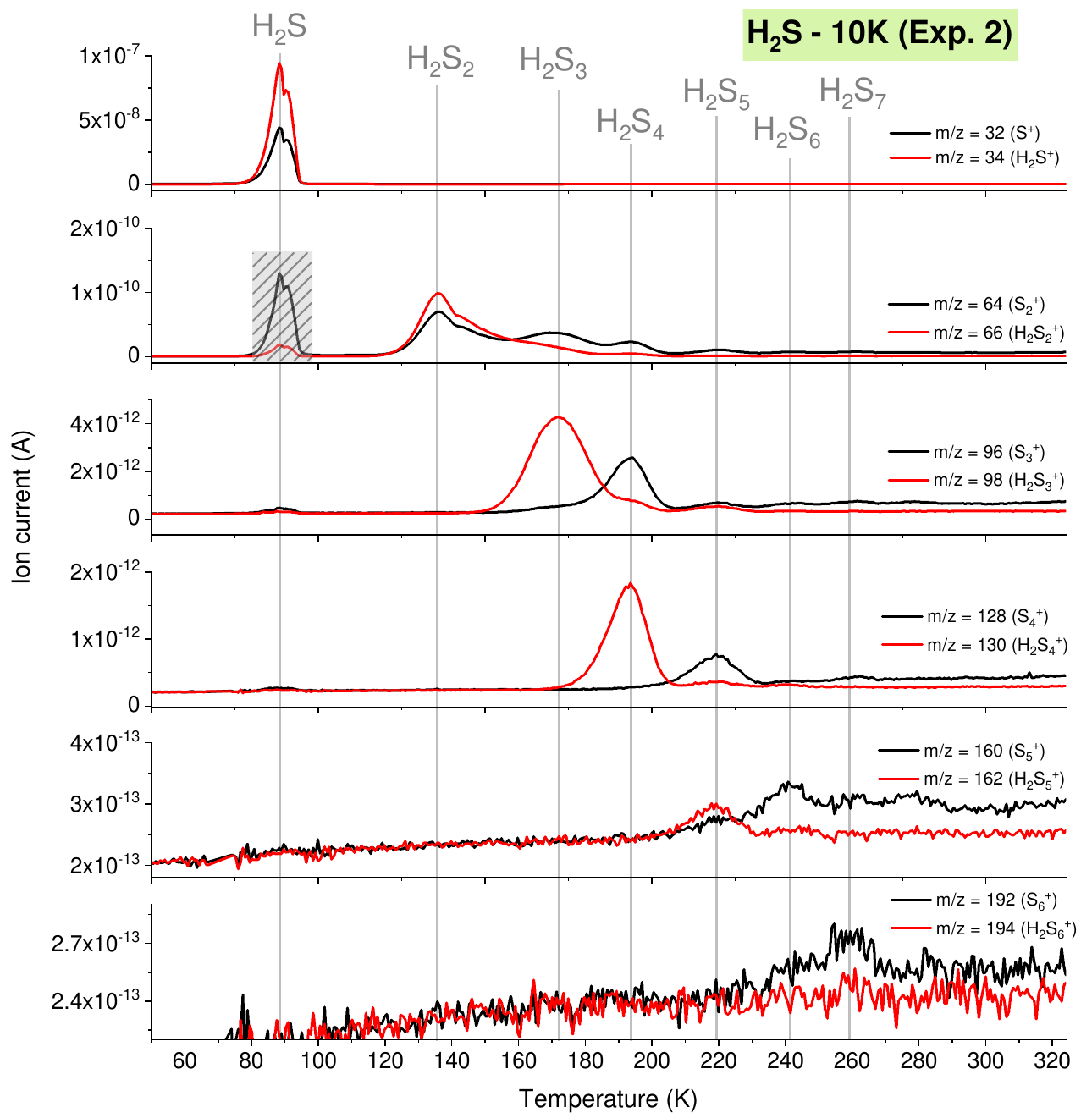}
  \caption{QMS data recorded during thermal desorption of H$_2$S ice sample deposited and irradiated at 10~K and irradiated at the same temperature (Exp. \textbf{2}). $\frac{m}{z}$ fragments related to S$_x$ and H$_2$S$_x$, x~=~1-6 are presented. Vertical lines indicate desorbing species at specific temperatures. Gray area corresponds to contamination coming from deposition, which was also observed in blank experiment, see Sect. \ref{sect.pure_H2S_ices}.}
  \label{Fig.desorcion_H2Sx_H2S_10K}
\end{figure*}

Fig. \ref{Fig.desorcion_H2Sx_H2S_10K} shows the QMS data obtained during thermal desorption of an irradiated pure H$_2$S ice sample (Exp.~\textbf{2}). Each panel is related to thermal desorption of a given H$_2$S$_x$ species, with x~=~1-6. Thermal desorption of H$_2$S was detected at 88~K. $\frac{m}{z}~=~64$ and $\frac{m}{z}~=~66$ are also detected co-desorbing with H$_2$S, but blank experiments confirmed that the observed signals are mostly due to contamination of S$_2$ ($^{32}$S$_2$ and $^{32}$S$^{34}$S) and a small fraction of H$_2$S$_2$ during deposition of the ice sample. The recorded intensity for $\frac{m}{z}~=~64$ and $\frac{m}{z}~=~66$ during warming up of H$_2$S in Exp.~\textbf{2} was attributed to the same contamination. Same behaviour was observed for H$_2$S ice samples deposited and irradiated at 50~K (Exp.~\textbf{3}, data not shown).\\


Thermal desorption of H$_2$S$_2$, H$_2$S$_3$ and H$_2$S$_4$ (panels 2, 3 and 4 in Fig. \ref{Fig.desorcion_H2Sx_H2S_10K}) take place at 136~K, 172~K, and 194~K, in well agreement with those reported by \cite{cazaux2022_H2S_chains}. Furthermore, the larger H$_2$S column density used in these experiments compared to \cite{cazaux2022_H2S_chains} allows to detect the presence of larger and less abundant H$_2$S$_x$ species. Third panel in Fig.~\ref{Fig.desorcion_H2Sx_H2S_10K} shows thermal desorption of $\frac{m}{z}~=~96$ at 194~K. It is related to S$_3^+$ cation. S$_3^+$ could be produced from direct ionization of S$_3$ molecules in the QMS. However, by looking at the mass spectra, it is clear that $\frac{m}{z}~=~96$ is co-desorbing with $\frac{m}{z}~=~130$, shown in panel 4, suggesting that S$_3^+$ is a fragment from the parent H$_2$S$_4$ molecule. The same behaviour was found for H$_2$S$_5$ $\left( \frac{m}{z} = 162 \right)$ and S$_4^+$ fragment $\left( \frac{m}{z} = 128 \right)$ at 219~K. Therefore, it is expected that larger species would show a similar fragmentation pattern, allowing the detection of H$_2$S$_6$ (241 K) and H$_2$S$_7$ (259 K) by the corresponding S$_5^+$ $\left( \frac{m}{z} = 160 \right)$ and S$_6^+$ $\left( \frac{m}{z} = 192 \right)$ fragments present in panels 6 and 7 in Fig. \ref{Fig.desorcion_H2Sx_H2S_10K}.\\

\begin{figure}
  \centering 
  \includegraphics[width=0.45\textwidth]{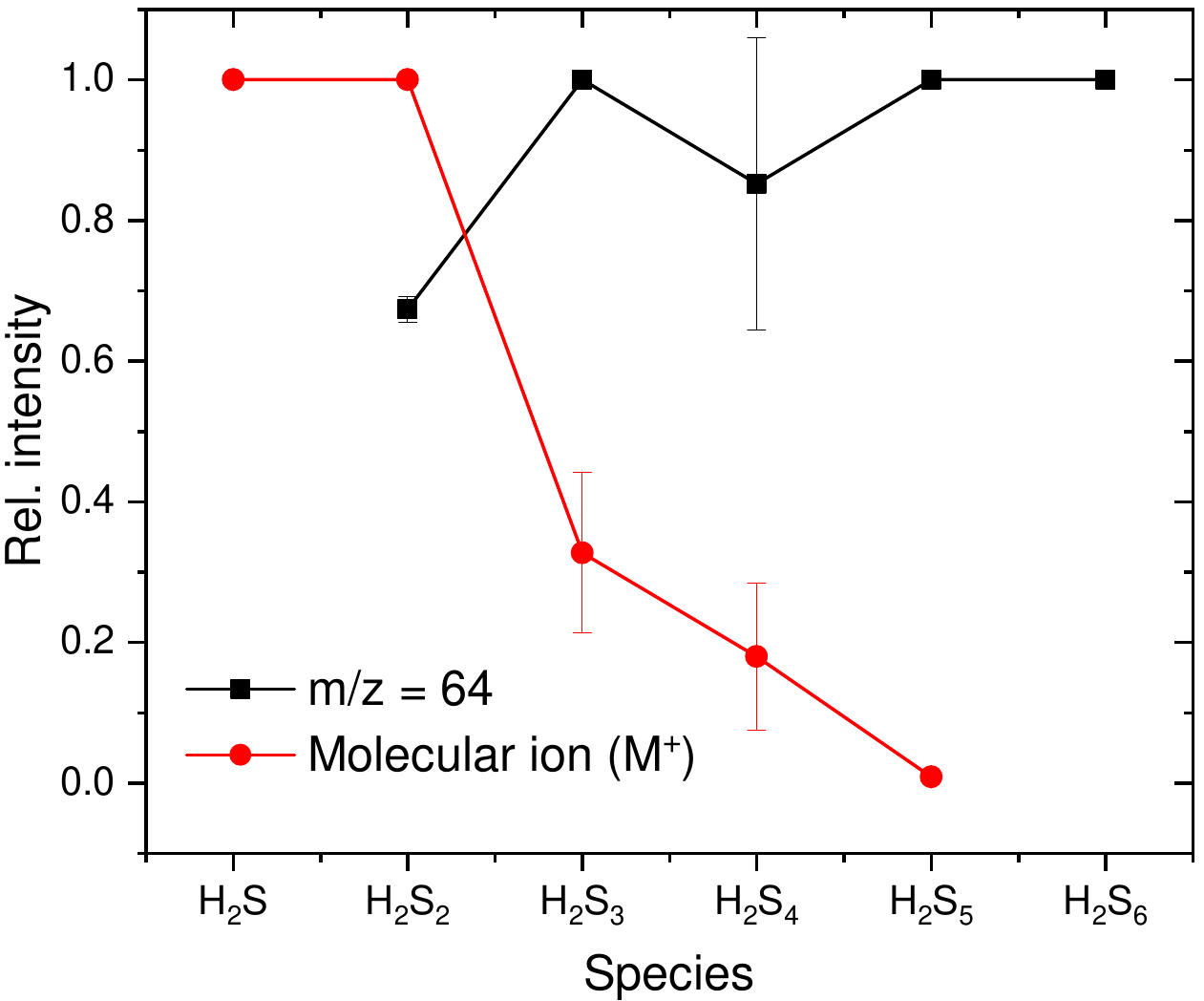}
  \caption{Relative intensity of $\frac{m}{z}$ fragment 64, and molecular ion (M$^+$) fragment for H$_2$S$_x$ species up to H$_2$S$_6$ in the mass spectra of each species (from Exps. \textbf{2} and \textbf{3}). As the number of S atoms in the molecule becomes larger, molecular ion fragment is less significant, while $\frac{m}{z}~=~64$ becomes the most intense feature in the mass spectra (see Fig.~\ref{Fig.mass_spectra_H2Sx}). Errors were obtained from the standard deviation of Exp. \textbf{2} and \textbf{3}.}
  \label{Fig.masa64_vs_masamolecular}
\end{figure}

Fig. \ref{Fig.masa64_vs_masamolecular} shows the normalized intensity of the molecular ion, M$^{+}$, for H$_2$S$_x$, x~=~1-6, and $\frac{m}{z}~=~64$, measured at their thermal desorption maximum during TPD. The decrease in the M$^+$ is indicative of the fragmentation of the species in the QMS. On the other hand, the increase for $\frac{m}{z}~=~64$ fragment, which becomes most prominent in the mass spectra for H$_2$S$_x$, x$\geq$3, suggests that $\frac{m}{z}~=~64$ fragment will still dominate for larger species. If larger H$_2$S$_x$ species are formed, the M$^+$ may not be intense enough to be detectable (as it is the case of H$_2$S$_6$ and H$_2$S$_7$ in Fig. \ref{Fig.desorcion_H2Sx_H2S_10K}), but $\frac{m}{z}~=~64$ may still be observable.\\

\begin{figure}
  \centering 
  \includegraphics[width=0.45\textwidth]{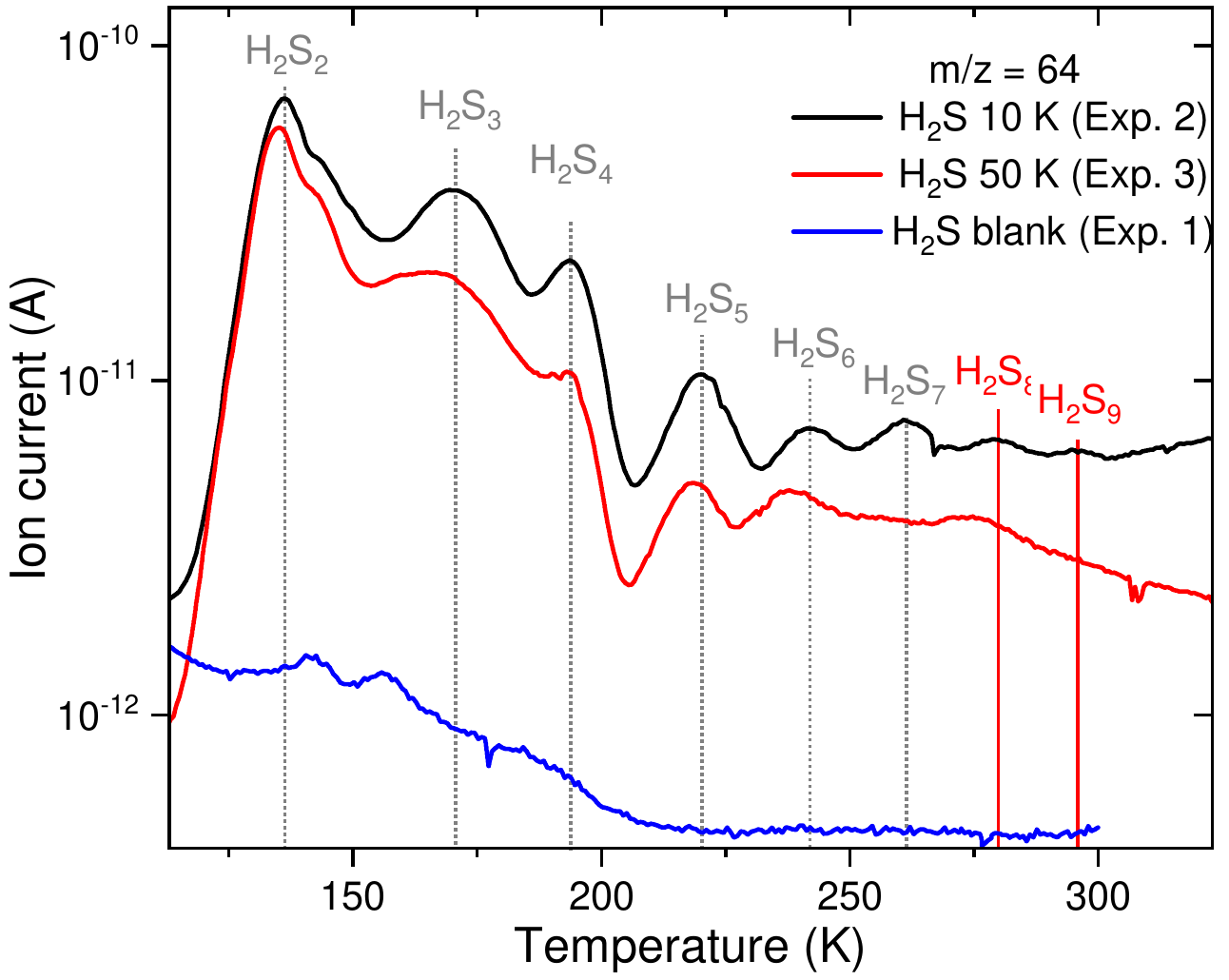}
  \caption{$\frac{m}{z}~=~64$ recorded for Exps. \textbf{1, 2, 3} during warm-up of the ice samples. Fig. \ref{Fig.mass_spectra_H2Sx} suggests that $\frac{m}{z}~=~64$ is the most intense mass fragment for H$_2$S$_x$ species when x is large enough. Therefore, the two desorption peaks located at 280~K and 296~K could be attributed to thermal desorption of H$_2$S$_8$ and H$_2$S$_9$.}
  \label{Fig.masa_64}
\end{figure}

Fig. \ref{Fig.masa_64} shows the thermal desorption profile of $\frac{m}{z}~=~64$ in Exps. \textbf{1}, \textbf{2}, and \textbf{3}. This figure clearly shows desorption peaks at different temperatures, which are coincident with thermal desorption of the H$_2$S$_x$ species up to H$_2$S$_7$. Furthermore, another two thermal desorption peaks are detected, which may be attributed to H$_2$S$_8$ and H$_2$S$_9$ molecules.\\

\begin{figure}
  \centering 
  \includegraphics[width=0.45\textwidth]{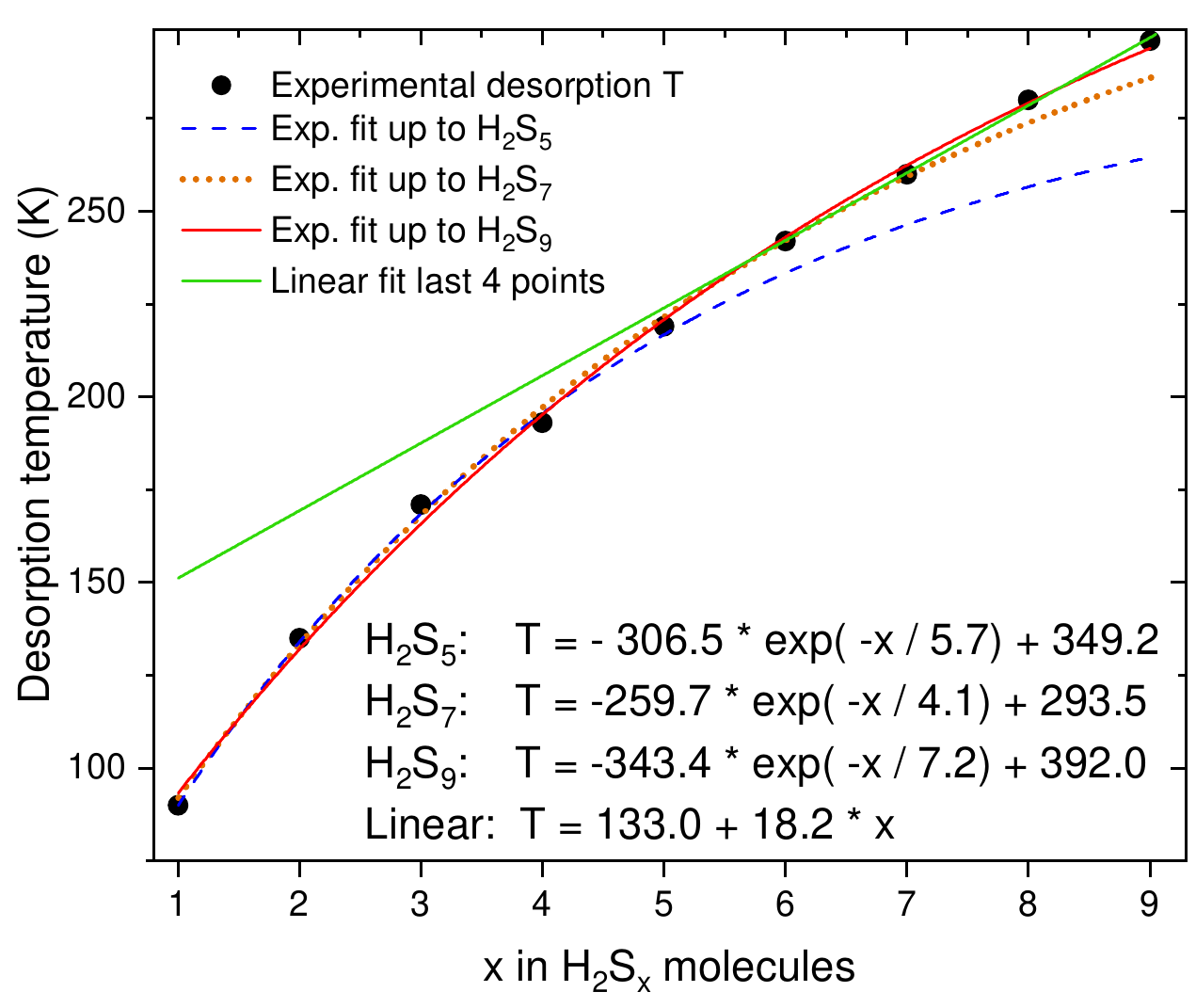}
  \caption{Experimental thermal desorption temperature for H$_2$S$_x$ species from Exp. \textbf{2}. Increasing number of S atoms should increase desorption temperature following an exponential trend (red solid line). Species up to H$_2$S$_9$ were detected experimentally.}
  \label{Fig.Desorption_temperature_H2Sx_fit}
\end{figure}

Another piece of evidence supporting the formation of these large polysulfides is related to their thermal desorption temperature. Experimental desorption temperatures confirm that each sulfur atom added to the chain will increase the thermal desorption temperature, but its relative significance in the species will decrease as the chain becomes longer. Therefore, an exponential fit is proposed to adjust thermal desorption temperature of H$_2$S$_x$ species up to H$_2$S$_5$ (blue dashed line in Fig. \ref{Fig.Desorption_temperature_H2Sx_fit}). This exponential fit should then predict thermal desorption temperature for larger species. As shown in Fig. \ref{Fig.Desorption_temperature_H2Sx_fit}, this fit understimates thermal desorption temperature for H$_2$S$_6$ and H$_2$S$_7$. Nevertheless, adding experimental desorption temperatures of H$_2$S$_6$ and H$_2$S$_7$ (orange dotted line), it also results in a good exponential fit.\\

If H$_2$S$_8$ and H$_2$S$_9$ species are responsible for the desorption peaks at 280~K and 296~K, their thermal desorption temperature should be well predicted by the exponential fit in Fig. \ref{Fig.Desorption_temperature_H2Sx_fit}. As it happened for H$_2$S$_6$ and H$_2$S$_7$, the exponential fit underestimates the desorption temperature for H$_2$S$_8$ and H$_2$S$_9$. The repeated underestimation of desorption temperatures suggests that the exponential fit is not reproducing well the experimental data for long polysulfides. \cite{Pouillot_1996_sublimacion_alcanos} studied the sublimation temperatures for alkanes from C$_{20}$H$_{42}$ to C$_{35}$H$_{72}$. They fitted thermal desorption temperatures using a linear fit. Green solid line in Fig. \ref{Fig.Desorption_temperature_H2Sx_fit} shows a linear fit of the last 4 experimental points, related to H$_2$S$_5$ - H$_2$S$_9$. Looking at the graph, it seems that the linear fit is adequate when dealing with relatively large species, but it fails for smaller sulfur chains, for which an exponential trend is more consistent with experimental data. Therefore we consider that this fit would make a better prediction for even longer H$_2$S$_x$ species.\\
 
\begin{figure*}
  \centering 
  \includegraphics[width=0.78\textwidth]{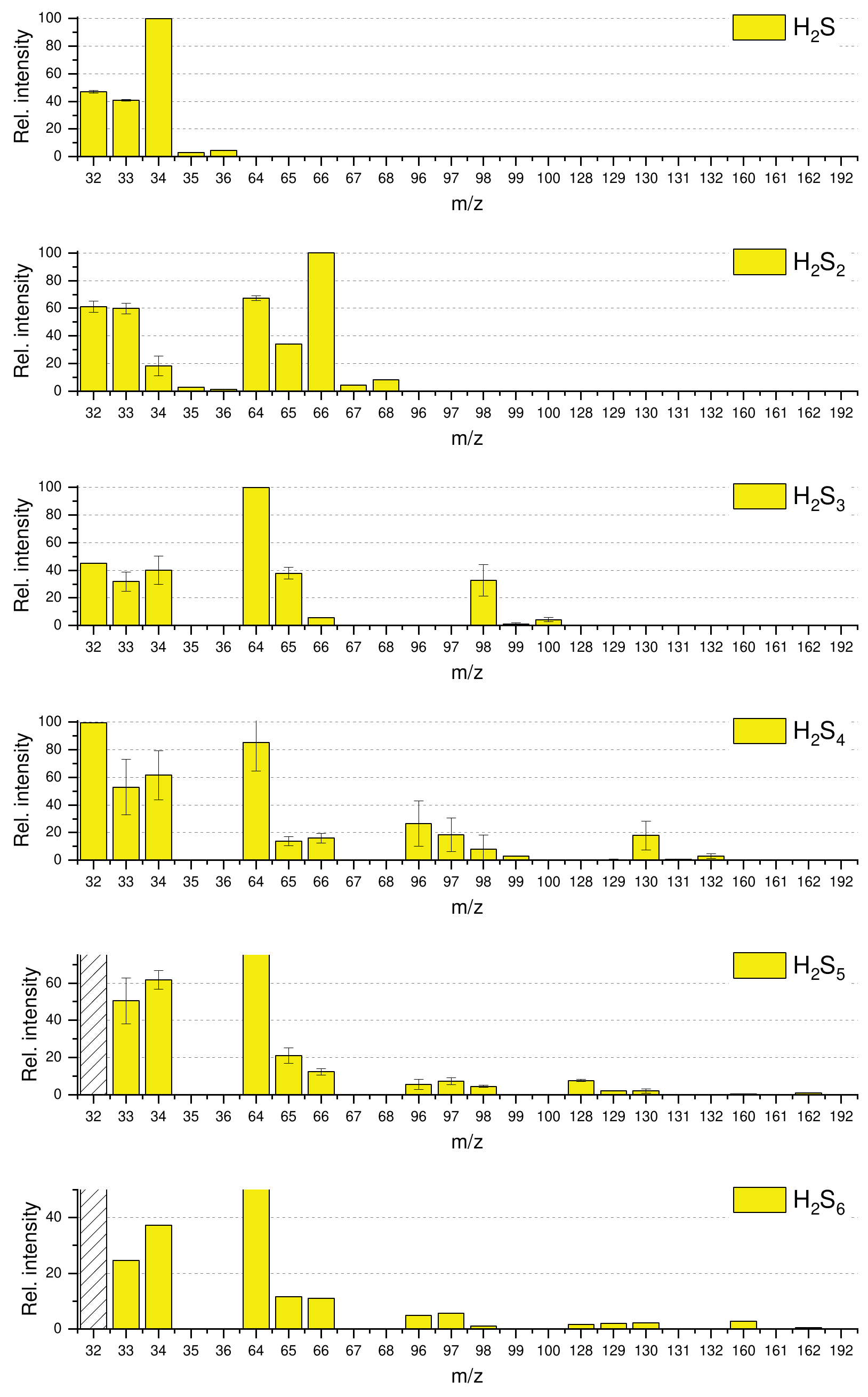}
  \caption{Experimental mass spectra of H$_2$S$_x$ (x~=~1-6) molecules obtained during thermal desorption of each species in Exps. \textbf{2}, \textbf{3}. $\frac{m}{z}~=~64$ was truncated for H$_2$S$_5$ and H$_2$S$_6$ to better show the contribution of minor fragments. Intensity of $\frac{m}{z}~=~32$ could not be determined for H$_2$S$_5$ and H$_2$S$_6$ due to the low abundance of these species, and the relatively high baseline of $\frac{m}{z}~=~32$. Some error bars are too small to be appreciated in the graph. Error bars for H$_2$S$_6$ could not be calculated, as there was not enough signal in Exp. \textbf{3} to obtain quantitative results for this species.}  \label{Fig.mass_spectra_H2Sx}
\end{figure*}

Fig. \ref{Fig.mass_spectra_H2Sx} and Table \ref{Table.mass_spectra} present the mass spectra of H$_2$S$_x$ molecules up to x~=~6, calculated from laboratory experiments during warming up of photoprocessed pure H$_2$S ice samples (Exps. \textbf{2}, \textbf{3}). The molecular ion (M$^+$) is not the most intense peak for those chains containing 3 or more sulfur atoms. Instead, $\frac{m}{z}~=~64$ becomes more intense for longer H$_2$S$_x$ species. It is important to note that, for all sulfur chains, the $\frac{m}{z}~=~M-34$ is relatively intense. Looking at the mass spectra at a specific temperature, the detection of M-34 could be wrongly attributed to the presence of the corresponding S$_{x-1}$ molecule, but it would arise from the characteristic fragmentation of H$_2$S$_x$.\\

\begin{table*} 
\scriptsize
    \caption[]{Mass spectra recorded during thermal desorption of H$_2$S$_x$ species in Exp. \textbf{2} and Exp. \textbf{3} (averaged values). Intensity was normalized to 100 for the most intense $\frac{m}{z}$ ratio of each species.}
    \label{Table.mass_spectra}
    \begin{tabular}{ccccccccccccccccccccccccc}
\hline
Species&32&33&34&35&36&64&65&66&67&68&96&97&98&99&100&128&129&130&131&132&160&161&162&192\\

\hline
\noalign{\smallskip}
\rowcolor{yellow!50}
H$_2$S&46&40&100&3&4&-&-&-&-&-&-&-&-&-&-&-&-&-&-&-&-&-&-&- \\
\noalign{\smallskip}
H$_2$S$_2$&58&57&13&-&-&69&34&100&4&8&-&-&-&-&-&-&-&-&-&-&-&-&-&-\\
\noalign{\smallskip}
\rowcolor{yellow!50}
H$_2$S$_3$&45&37&47&-&-&100&35&6&-&-&0.1&0.6&25&0.5&3&-&-&-&-&-&-&-&-&-\\
\noalign{\smallskip}
H$_2$S$_4$&100&39&49&-&-&70&11&14&-&-&15&10&0.3&-&-&-&0.2&10&0.4&2&-&-&-&-\\
\noalign{\smallskip}
\rowcolor{yellow!50}
H$_2$S$_5$&?&42&58&-&-&100&18&11&-&-&4&6&4&-&-&8&2&1&-&-&0.2&-&0.9&-\\
\noalign{\smallskip}
H$_2$S$_6$&?&25&37&-&-&100&12&11&-&-&5&6&1&-&-&2&2&2&-&-&3&-&0.5&0.2\\
\noalign{\smallskip}
\hline
\end{tabular}\\
\scriptsize{}
\end{table*}

Table \ref{Table.isotopos} shows the thermal desorption temperature measured for H$_2$S$_x$ species up to H$_2$S$_9$. Last column shows the probability of having one $^{34}$S atom in the molecule. As the chains become larger, the fraction of molecules containing the $^{34}$S isotope is higher. Therefore, the detection of $\frac{m}{z}~=~$M+2 is only indicative of the presence of a different species if the expected $\frac{(M+2)}{M}$ ratio is significantly larger than the statistical probability (last column in Table \ref{Table.isotopos}). For Exps. \textbf{2} and \textbf{3}, the $\frac{(M+2)}{M}$ ratio was calculated. Results are shown in column 3 of Table \ref{Table.isotopos}. Roughly, the isotopic expected ratio was obtained for H$_2$S$_x$, x~=~1-4, meaning that species with H$_4$S$_x$ as molecular formula are not formed, and the presence of M+2 mass fragment is only due to isotopic contribution. For larger polysulfides, the recorded signal of M+2 mass-to-charge fragment was not intense enough to calculate the $\frac{(M+2)}{M}$ ratio.\\

\begin{table} 
   \centering
    \caption[]{Second column: thermal desorption temperature of H$_2$S$_x$ species measured in Exps. \textbf{2}, \textbf{3}. Third column: ratio between the ion current measured by the QMS during thermal desorption of H$_2$S$_x$ species for molecular mass-to-charge ratio (M) and M+2, mostly due to isotopic contribution of $^{34}$S atoms. For these ratios, errors are estimated to be about 10\%. The low intensity recorded for the corresponding M+2 mass-to-charge in H$_2$S$_x$ species with x > 4 prevented us from calculation of this ratio. Last column shows the probability of $^{34}$S in these molecules.}
    \label{Table.isotopos}
    \begin{tabular}{cccc}
\hline
S atoms& T$_{des}$ (K) & Ratio (M+2)/M & Probability (\%) \\
\hline
\noalign{\smallskip}
\rowcolor{yellow!50}
1&88   &4.4 &4.2\\
\noalign{\smallskip}
2&136  &8.1 &8.3\\
\noalign{\smallskip}
\rowcolor{yellow!50}
3&172  &13.1 &12.2\\
\noalign{\smallskip}
4&194  &16.5 &15.9\\
\noalign{\smallskip}
\rowcolor{yellow!50}
5&219   &-&19.5\\
\noalign{\smallskip}
6&241   &-&22.9\\
\noalign{\smallskip}
\rowcolor{yellow!50}
7&259   &-&26.2\\
\noalign{\smallskip}
8&280   &-&29.3\\
\noalign{\smallskip}
\rowcolor{yellow!50}
9&296   &-&32.3\\
\noalign{\smallskip}
\hline
\end{tabular}\\
\scriptsize{}
\end{table}

\subsection{Water effect on the formation and thermal desorption of sulfur chains}
\label{sect.H2O_H2S_ices}
\begin{figure}
  \centering 
  \includegraphics[width=0.5\textwidth]{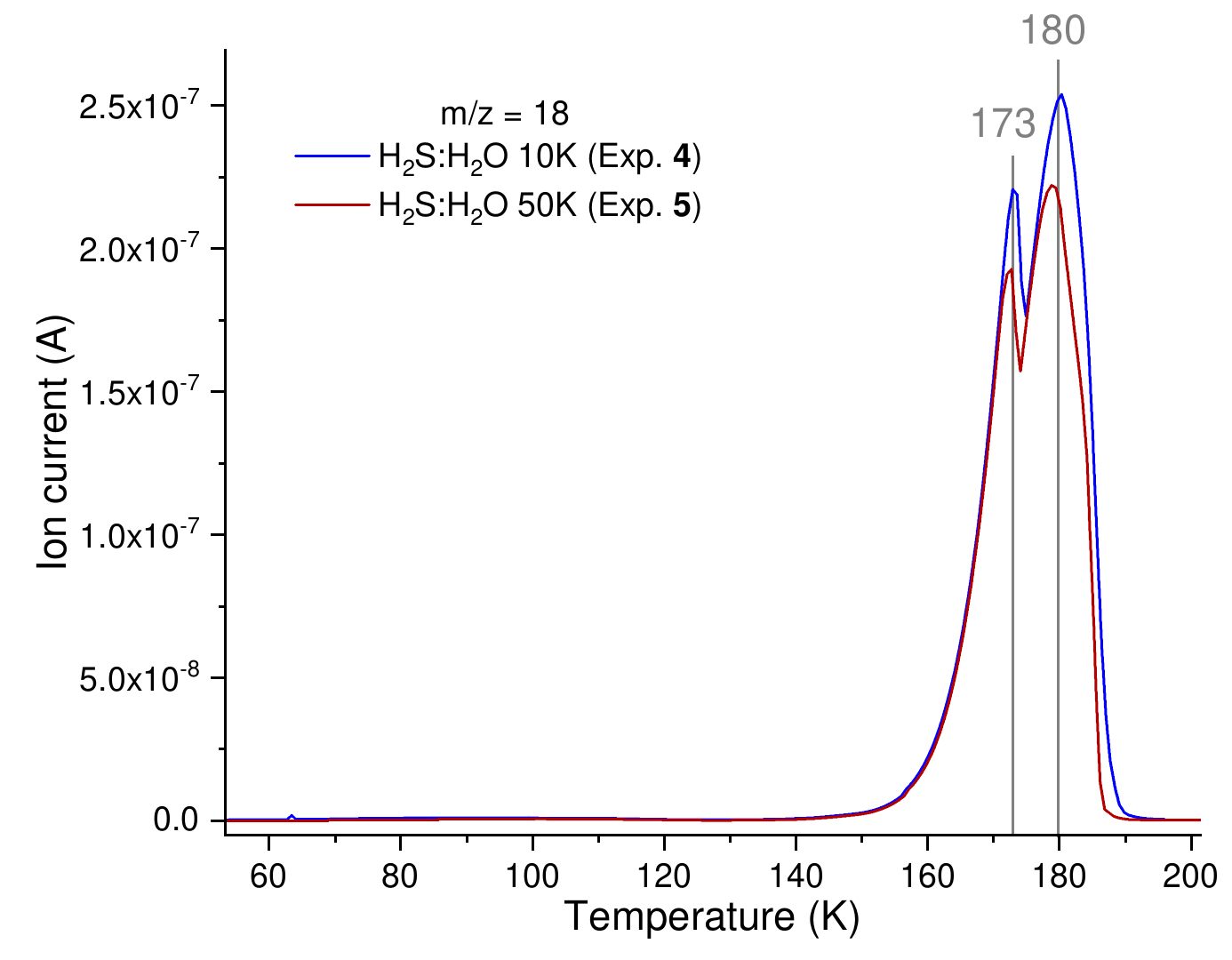}
  \caption{Thermal desorption of water molecules from Exps. \textbf{4}, \textbf{5}. The two desorption peaks could be related to desorption of different crystalline phases of water (i. e. cubic and hexagonal water ice).}
  \label{Fig.desorcion_H2O_comparison}
\end{figure}

\begin{figure*}
  \centering 
  \includegraphics[width=\textwidth]{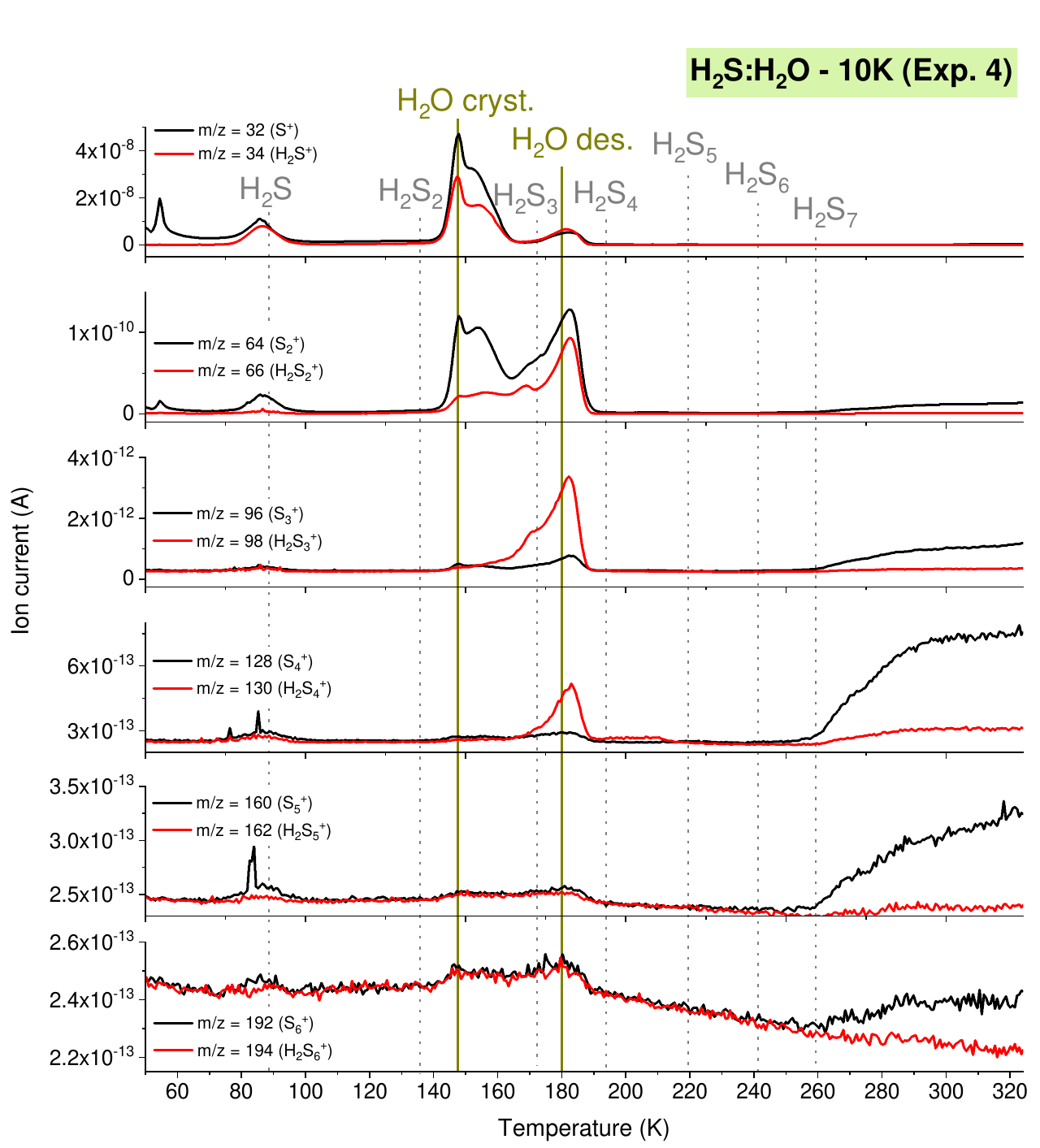}
  \caption{QMS data recorded during TPD of Exp.~\textbf{4}. $\frac{m}{z}$ fragments related to S$_x^+$ and H$_2$S$_x^+$, x~=~1-6, ions are shown (same $\frac{m}{z}$ fragments shown in Fig. \ref{Fig.desorcion_H2Sx_H2S_10K}). Vertical dotted lines correspond to thermal desorption of H$_2$S$_x$ species in Exps. \textbf{2}, \textbf{3}. Vertical brownish green lines are related to H$_2$O crystallization temperature and H$_2$O thermal desorption. Some sulfur species are ejected from the ice during water transition from amorphous to cubic crystalline ice \citep[volacno desorption, ][]{Smith1997_volcano_desorption_water, Martin_domenech2014_TPD_hielos}. All sulfur species codesorb in the water ice matrix at 180~K.}
  \label{Fig.desorcion__H2Sx_H2S_H2O_10K}
\end{figure*}

Fig.~\ref{Fig.desorcion_H2O_comparison} shows thermal desorption of water molecules in Exps. \textbf{4} and \textbf{5}. Two peaks are observed at 173~K and 180~K, probably due to thermal desorption of two different water phases. Below 150~K no thermal desorption is observed. Therefore, the thermal desorption observed around 140~K for H$_2$S$_x$ species in these experiments is not due to co-desorption with water molecules or H$_2$S (Fig.~\ref{Fig.desorcion__H2Sx_H2S_H2O_10K}). According to \cite{Smith1997_volcano_desorption_water}, this desorption is induced by the water transition from amorphous to cubic crystalline structure, ejecting molecules during the process. This effect is known as volcano desorption.\\

\begin{figure*}
  \centering 
  \includegraphics[width=\textwidth]{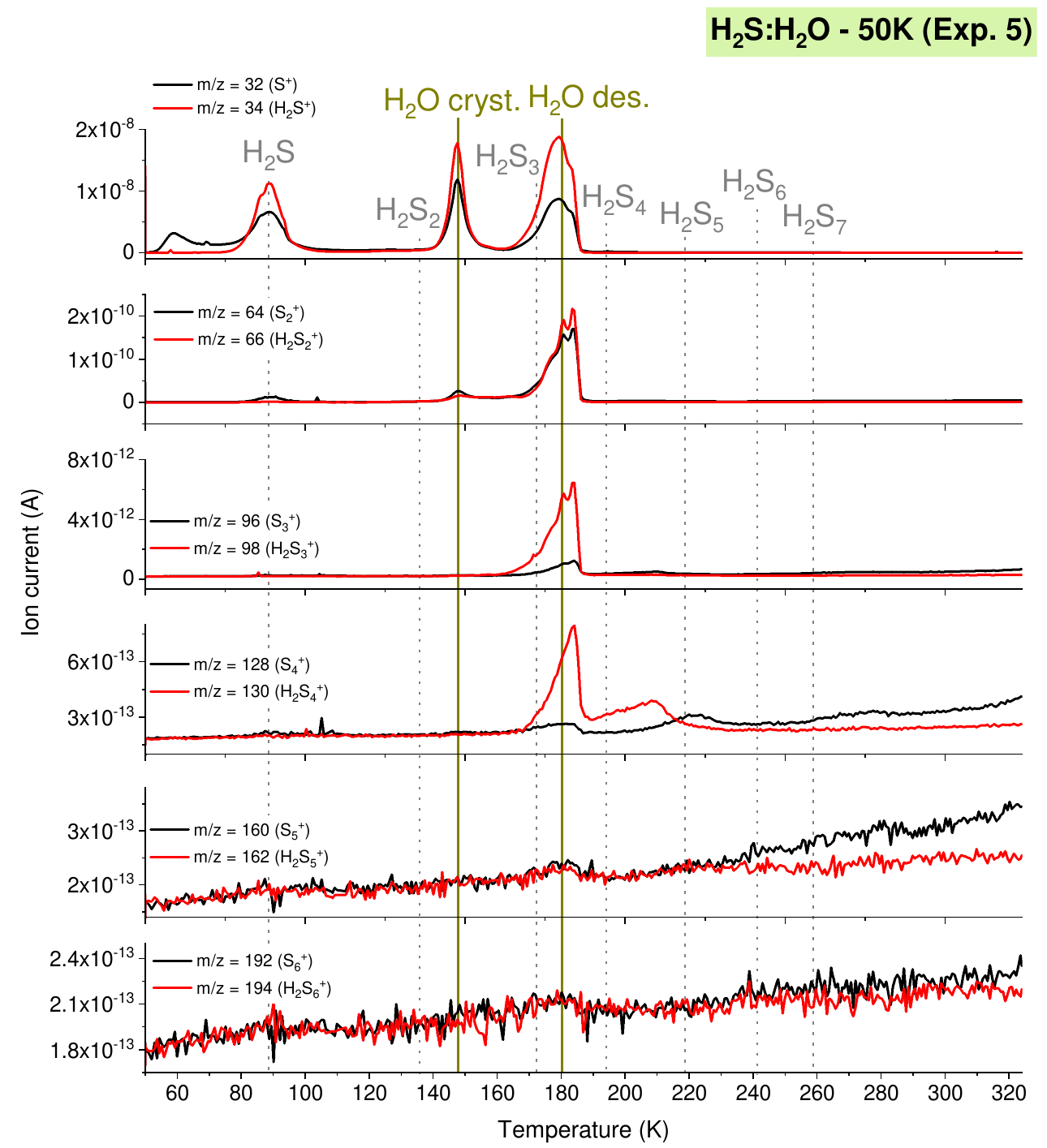}
  \caption{QMS data recorded during TPD of Exp.~\textbf{5}. $\frac{m}{z}$ fragments related to S$_x^+$ and H$_2$S$_x^+$, x~=~1-6, ions are shown (same $\frac{m}{z}$ fragments shown in Fig. \ref{Fig.desorcion_H2Sx_H2S_10K}). Vertical dotted lines correspond to thermal desorption of H$_2$S$_x$ species in Exps. \textbf{2}, \textbf{3}. Vertical dark yellow lines are related to H$_2$O crystallization temperature and H$_2$O thermal desorption. Desorption of H$_2$S$_x$ in the lower pannels is less intense than the one recorded for Exp. \textbf{4} (Fig. \ref{Fig.desorcion__H2Sx_H2S_H2O_10K}), indicative of the lower formation of long sulfur chains at 50~K, see text.}
  \label{Fig.desorcion__H2Sx_H2S_H2O_50K}
\end{figure*}

Fig.~\ref{Fig.desorcion__H2Sx_H2S_H2O_10K} shows thermal desorption of polysulfides (H$_2$S to H$_2$S$_7$) in H$_2$O-rich ice samples in Exp. \textbf{4}. This figure shows a relatively small thermal desorption of H$_2$S molecules at the expected temperature from pure H$_2$S ice (87~K), and a larger volcano desorption during water transition from the amorphous to cubic crystalline structure. As explained in Sect. \ref{sect.pure_H2S_ices}, the $\frac{m}{z} = 64$ signal measured at 88~K is related to contamination, as it happenned in Exp. \textbf{2}. The larger intensity of $\frac{m}{z} = 64$ recorded during H$_2$O crystallization could be, in principle, also due to contamination. If this is the case, a similar intensity should be recorded in Exp. \textbf{5}. Fig. \ref{Fig.desorcion__H2Sx_H2S_H2O_50K} shows almost no $\frac{m}{z} = 64$ signal at 147~K, negligible in comparison with the signal measured for the 10~K experiment. It can be concluded that S$_2$ is really formed in the 10~K experiment, but it is not observed in the 50~K experiment. The absence of S atoms and S$_2$ molecules is indicative of the higher mobility of species at 50~K, determining a larger reactivity of radicals as they are formed, opposite to the accumulative effect described previously for the low temperature experiments. H$_2$S$_x$ species show a similar behaviour with respect to the 10~K experiment, although some differences regarding their intensity are highlighted hereafter.\\

Furthermore, the intensity recorded for $\frac{m}{z} = 32$ during thermal desorption of H$_2$S in Exp. \textbf{4} (Fig. \ref{Fig.desorcion__H2Sx_H2S_H2O_10K}) is larger than the intensity expected during thermal desorption of H$_2$S (Fig. \ref{Fig.mass_spectra_H2Sx}). Therefore, there is another source of S$^+$, intense enough to produce such a large difference in the $\frac{m}{z} = 32$ intensity. The source of S$^+$ could be: 1) fragmentation of S$_2$ molecules, which seem to be abundant in Exp. \textbf{4}, or 2) desorption of S atoms trapped in the water matrix. Our data does not allow us to distinguish both scenarios. However, it can be concluded that during water transition around 148~K, most of the S atoms and S$_2$ molecules, together with a large fraction of H$_2$S, would be expelled from the bulk of the ice.

All the H$_2$S$_x$ species show a small volcano desorption at 148~K during water transition, and a more intense thermal desorption during water desorption at 180~K. Most of the polysulfides are dragged with H$_2$O molecules during H$_2$O thermal desorption in Exps. \textbf{4} and \textbf{5}. Interestingly, a relatively small fraction of H$_2$S$_4$ and H$_2$S$_5$ were detected in the gas phase (detected by $\frac{m}{z} = 130$ and $\frac{m}{z} = 128$, respectively) at 208~K and 222~K, close to their pure thermal desorption temperature (194~K and 219~K, Table \ref{Table.isotopos}), indicative of a small thermal desorption of the less volatile polysulfides apart from H$_2$O desorption.\\

From 260~K, Fig. \ref{Fig.desorcion__H2Sx_H2S_H2O_10K} shows an increase in the intensity of mass-to-charge ratios related to S$_x^+$ cations. The corresponding H$_2$S$_x$ mass-to-charge ratios are, however, almost not altered at this temperature, suggesting that non-hydrogenated species are responsible of this effect. On Earth, S$_8$ is the most stable sulfur allotrope, and previous works have highlighted the efficient formation of S$_8$ in interstellar ice analogs \citep{Tesis_guillermo}. According to NIST database, mass spectra of S$_8$ produces S$_x^+$ cations, with x~=~2-8. S$_7$ and S$_8$ are beyond the limits of our QMS, but the simultaneous increase in the intensity of $\frac{m}{z} = 64, 96, 128, 160$, and $192$ suggest that S$_8$ could be responsible of this behaviour. \cite{Ferreira_2011_sublimation_sulfur} calculated the low-pressure phase diagram of sulfur, including experimental studies from \cite{bradley1951_sublimation_rhombic_sulphur} and \cite{briske1960_sulfur_sublimation}. The last two works made experiments on the vapour pressure of S$_8$ as a function of pressure. At the relatively low densities present in the ISM (and reproduced in our chamber), vapour pressure of solid-phase sulfur can induce sublimation from 260~K. A similar effect arises in the 50~K experiment (Fig. \ref{Fig.desorcion__H2Sx_H2S_H2O_50K}, Exp. \textbf{5}). However, the intensity recorded is much lower, suggesting that S$_8$ is more efficiently formed in the 10~K experiment.\\

After UV irradiation of H$_2$S:H$_2$O ice mixtures, no S-O bonds are detected in the IR spectrum, meaning that the formation of SO or SO$_2$ is not as favoured as the formation of sulfur chains, even though water was 4 times more abundant than H$_2$S molecules. This effect reveals that chemical reactions between sulfur species prevail over sulfur species reacting with water molecules or OH$\cdot$ radicals, as observed by \cite{Jimenez-escobar2011AA...536A..91J}.\\

\subsection{Temperature effect}
\label{sect.temperature_effect}

\begin{figure}
  \centering 
  \includegraphics[width=0.45\textwidth]{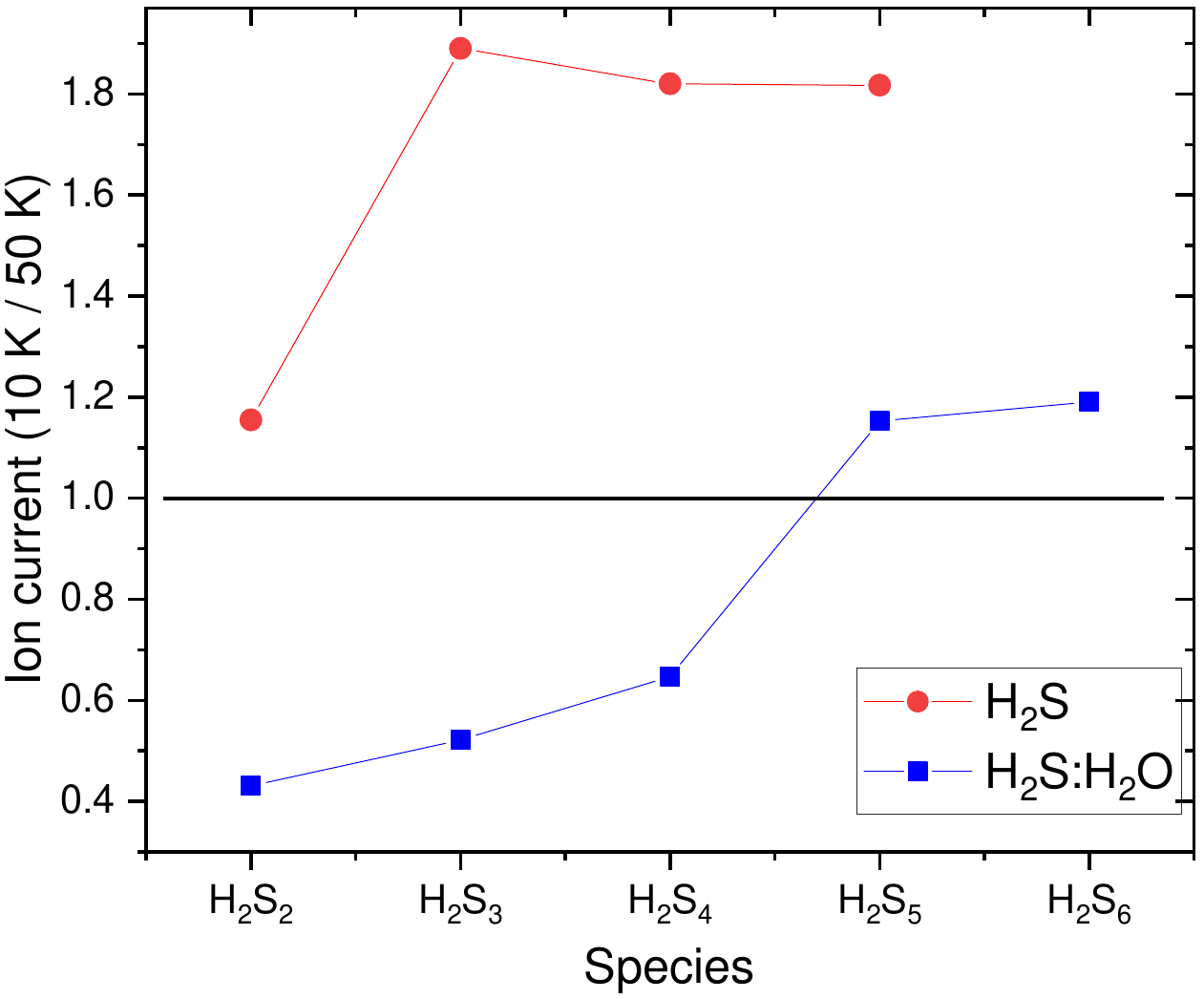}
  \caption{Ratio between the QMS signal recorded for experiments at 10~K (Exps.~\textbf{2}, \textbf{4}) and experiments performed at 50~K (Exps. \textbf{3}, \textbf{5}) during thermal desorption of each of the H$_2$S$_x$ species. While absolute quantification of the abundance of each species requires non-available data from each molecule \citep[see][]{MD2015}, the ratio between the abundance of a given species can be calculated directly from the QMS signal using the molecular fragment, M. Larger H$_2$S$_x$ species have a molecular mass larger than 200 amu, which is the largest mass-to-charge ratio that can be measured by our QMS.}
  \label{Fig.Ratio_desorcion_H2S_en_H2S-H2O}
\end{figure}

Fig.~\ref{Fig.Ratio_desorcion_H2S_en_H2S-H2O} shows the ratio between the ion current measured for the molecular ion during thermal desorption of H$_2$S$_x$ in the low temperature and high temperature experiments, both H$_2$S pure and H$_2$S:H$_2$O ice mixtures. A ratio greater than 1 is therefore indicative of the larger formation of a given species in the 10~K experiments. Regarding experiments with pure H$_2$S ice samples, the formation of H$_2$S$_x$ species is favoured at 10~K, specially long chains (x$\geq$3), for which the ion current measured during thermal desorption is almost double of the one measured for the low temperature experiment. On the other hand, H$_2$S:H$_2$O ice mixtures showed a larger formation of short sulfur chains (x$\leq$4) at 50~K, while the formation of long sulfur chains was almost the same at both temperatures.\\

A likely explanation of this effect is as follows: concerning pure H$_2$S ice samples, UV photons absorbed by H$_2$S molecules will mainly produce HS$\cdot$ radicals. At high temperature, as HS$\cdot$ radicals are produced, they will diffuse through the ice and react, forming H$_2$S$_2$, or relatively short chains, which are observed in similar abundances at both temperatures. At low temperature, as the mobility of these radicals is reduced and the porosity of the ice is increased, there is an accumulative effect of HS$\cdot$ along irradiation. During warming up of the ice samples, at a specific temperature, pores are expected to coalesce \citep{cazaux_2015_AA...573A..16C}, enabling radicals to diffuse and react, all of them at nearly the same time, favouring the formation of larger chains. As a result, for pure H$_2$S ice samples, the formation of long sulfur chains is almost doubled at 10~K. Indeed, H$_2$S$_x$, x$\geq$7, were not detected during the high temperature experiment (see Fig. \ref{Fig.masa_64}). Radicals produced from irradiation of H$_2$S:H$_2$O ice mixtures will also be more reactive at 50~K, what determines the higher abundance of short chains at 50~K. The accumulative effect of radicals in the 10~K experiment determines an increase in the 10~K/50~K ratio for the larger polysulfides.\\

Sulfur detection agrees with previous explanation. S$_8$ can be formed from successive addition of S atoms, or by dehydrogenation of H$_2$S$_x$ species. As shorter S$_x$ species were not detected, the first mechanism is less probable. If S$_8$ is produced from dehydrogenation of H$_2$S$_x$, the larger formation of large polysulfides at low temperature will favour the formation of S$_8$, which is, in fact, more abundant in the 10~K experiment.\\

\section{Comparison with Rosetta}
\label{sect.comparison_rosetta}


Experiments suggest that H$_2$S$_x$ species are formed both in pure H$_2$S ice and H$_2$S:H$_2$O ice mixtures. The detection of these species in the gas phase is, however, highly dependant on the environment. In water matrix, most of the photoproducts are co-desorbing following H$_2$O ice crystallization or H$_2$O thermal desorption. Surrounded by H$_2$S, each species sublimated at a specific termperature, shown in Table \ref{Table.isotopos}.\\

\cite{Calmonet2016_detection_S3_S4_Rosetta} reported the first detection of several sulfur bearing molecules in comet 67P/Churyumov-Gerasimenko, as measured with the ROSINA mass spectrometer. In particular, they measured a peak related to signal $\frac{m}{z} = 96$, which was attributed to the S$_3^+$ cation. As the comet got closer to the perihelion, an extended mass range of the mass spectrometer revealed also the presence of $\frac{m}{z} = 128$, which was related to S$_4^+$ cation. However, as noted by the authors, the lack of experimental data prevented them from confirming the parent molecules of these cations. Fig. \ref{Fig.desorcion_H2Sx_H2S_10K}, Fig. \ref{Fig.mass_spectra_H2Sx} and Table \ref{Table.mass_spectra} suggest that H$_2$S$_4$ and H$_2$S$_5$ may be the parent molecules producing $\frac{m}{z} = 96, 128$. The molecular mass fragment for these species (mass 130 and 162, respectively) may have an intensity too low to be detected with ROSINA, leaving the M-34 fragments as the most characteristic mass-to-charge units, in line with the mass spectra measured in our experiments (see Fig. \ref{Fig.mass_spectra_H2Sx}).\\

As shown in Fig. \ref{Fig.desorcion_H2Sx_H2S_10K}, thermal desorption of polysulfides is, as expected, less abundant as the chain size increases. Therefore, if H$_2$S$_4$ and H$_2$S$_5$ are the molecules present in the mass spectra of ROSINA presented by \cite{Calmonet2016_detection_S3_S4_Rosetta}, H$_2$S$_2$ and H$_2$S$_3$ might be also detected. \cite{Calmonet2016_detection_S3_S4_Rosetta} reported upper limits for the detection of H$_2$S$_2$. However, as stated by the authors, the lack of experimental data for ionization cross section and fragmentation pattern for H$_2$S$_2$ induced a large error on these calculations. They assumed the fragmentation pattern of C$_2$H$_2$, which, according to NIST database and the mass spectra shown in Fig. \ref{Fig.mass_spectra_H2Sx}, are significantly different. This fact, combined with the large variety of species which may contribute to $\frac{m}{z}~=~66$, may hinder the detection and quantification of H$_2$S$_2$. The detection of H$_2$S$_3$ with ROSINA was recently reported by \cite{Mahjoub2023_sulfur_species_rosetta}, meaning that $\frac{m}{z}~=~98$ was also present in the mass spectra. Larger chains may be present in comet 67P, but an extended mass-to-charge range and better sensitivity might be needed to detect their presence.\\

However, results presented in Sect. \ref{Results} show another possibility. S$_8$ molecules may be sublimating from the ice structure. The fragmentation pattern of S$_8$ is mainly formed by S$_x^+$, 2~$\geq$~x~$\leq$~8, cations. The mass range of ROSINA (up to 150 amu) included S$_2^+$, S$_3^+$ and S$_4^+$ mass-to-charge fragments, but longer chains, even if present, would not have been detected. With this work, we propose two possibilities for the identification of the parent molecules in comet 67P: 1) H$_2$S$_4$ and H$_2$S$_5$ to be responsible of those S$_3^+$ and S$_4^+$ cations, respectively, or 2) S$_8$ species, sublimating and being fragmented in the mass spectrometer. If S$_8$ is the parent molecule, then S$_5^+$ and S$_6^+$ cations could be also detected if the mass spectrometer would have allowed detection at larger amu.\\

\section{Conclusions and astrophysical implications}
\label{sect.conclusions}

In this work we studied experimentally H$_2$S ice samples, as well as more realistic H$_2$S:H$_2$O ice mixtures, regarding the formation of polysulfides and sulfur allotropes. Most important conclusions extracted from this work are highlighted here:

\begin{itemize}

    \item Irradiation of H$_2$S ice samples led to thermal desorption of H$_2$S$_x$ species up to H$_2$S$_9$. Exponential and linear fits of these temperatures suggest that thermal desorption temperature of larger polysulfides can be predicted using the T$_{des} = 133 + 18.2$ $\cdot$~x linear fit, for H$_2$S$_x$ species. Consequently, large polysulfides may be present in the gas phase of star-forming regions and protoplanetary disks at the given thermal desorption temperatures (Table \ref{Table.isotopos}).\\
    
    \item In contrast, H$_2$O-rich ice samples showed thermal desorption of large polysulfides at two specific temperatures. At 148~K in laboratory experiments ($\approx$98~K in space), a volcano desorption was found, due to H$_2$O transition from amorphous to cubic crystalline ice. At 180~K in laboratory experiments ($\approx$120~K in space) there is a second desorption during H$_2$O pure thermal desorption, which drags most of the polysulfides remaining in the ice matrix. Only a small fraction of H$_2$S$_4$ and H$_2$S$_5$ were detected during thermal desorption at temperatures larger than H$_2$O desorption. Furthermore, the detection temperature of H$_2$S$_x$ could be indicative of the pure or mixed composition of the ice mantle.\\


    \item H$_2$S embedded in water molecules is able to form sulfur allotropes, which converge into the final formation of S$_8$. Although the vapour pressure of S$_8$ may induce its desorption at the low pressures present in the ISM, it is not expected to have S$_8$ in the gas phase. Sulfur sublimation measured from 260~K in our experiments, in contrast with S$_8$ detection in the solid residue at 300~K \citep[e. g.][]{Tesis_guillermo} reveal that S$_8$ can be kept in the solid phase or sublimate depending on the non-covalent interactions established. Heterogeneity of species tend to retain them in the residue, even if those species would sublimate in its pure form.\\

    \item The formation of sulfur chains is temperature dependent. Both in H$_2$S pure ice and H$_2$S:H$_2$O ice mixtures, there is a preferential formation of large polysulfides (H$_2$S$_x$) when deposition and irradiation take place at low (10 K) temperatures, while shorter species are favoured in high (50 K) irradiation experiments. This effect is explained by the different diffusion of radicals depending on the temperature (Sect. \ref{sect.temperature_effect}). Sublimating S$_8$ allotrope (detected in H$_2$S:H$_2$O ice mixtures) was also more abundant in the low temperature experiment (Figs. \ref{Fig.desorcion__H2Sx_H2S_H2O_10K} and \ref{Fig.desorcion__H2Sx_H2S_H2O_50K}).\\

    \item \cite{Calmonet2016_detection_S3_S4_Rosetta} detected S$_3^+$ and S$_4^+$ cations with ROSINA instrument on board on the Rosetta mission. The origin of these fragments could be: 1) S$_3$ and S$_4$ molecules themselves or 2) larger species fragmented in the mass spectrometer into S$_3^+$ and S$_4^+$ cations. The lack of experimental data and the mass range of ROSINA (from 1 to 150 amu) prevented them from confirming their origin. We propose H$_2$S$_4$ and H$_2$S$_5$ as possible parent molecules, which is in line with the H$_2$S$_3$ detection reported by \cite{Mahjoub2023_sulfur_species_rosetta}. S$_8$ could also be the parent molecule for these fragments, according to its sublimation observed for temperatures larger than 260~K in H$_2$S:H$_2$O ice mixtures. However, the heterogeneous nature of ice mantles suggests that S$_8$ could be kept in the refractory residue in the ISM, due to non-covalent interactions with more polar species.\\

    \item In the ISM, sulfur atoms may react in the translucent phase \citep{cazaux2022_H2S_chains}, thus forming sulfur chains, which would probably end up as the most stable S$_8$ allotrope. According to our results on ice processing, longer chains are favoured when H$_2$S photoprocessing takes place at low temperatures, while shorter chains would be formed if irradiation takes place at higher temperatures.\\

    \item Mass spectra of H$_2$S$_x$ species up to H$_2$S$_6$ have been measured for the first time (Fig. \ref{Fig.mass_spectra_H2Sx}). As the number of sulfur atoms is increased, the molecular mass-to-charge fragment intensity is lower, while the intensity of the $S_{x-1}$ mass-to-charge fragment increases, what is relevant for the detection of sulfur-based species in the ISM, hot cores, protoplanetary disks and comets. For future observations of sulfur, observations at mass-to-charge ratios of S$_x^+$ (32, 64, 96, 128, 160, 192...) and H$_2$S$_x^+$ cations (34, 66, 98, 130, 162, 194...) will allow to determine whether the observations can be linked to the presence of long sulfur chains, and distinguish whether polysulfides of sulfur allotropes are present by comparing with the mass spectra reported here. A mass range up to 256 amu will be required to detect the presence of S$_8$.\\ 

\end{itemize}

\section{Acknowledgements}
This research was funded by project PID2020-118974GB-C21 by the Spanish Ministry of Science and Innovation. The project leading to these results has received funding from "La Caixa" Foundation, under agreement LCF/BQ/PI22/11910030. This project is co-funded by the European Union (ERC) under ERC-AdG-2022 grant SUL4LIFE ( GA 101096293).

\section{Data availability}
The data underlying this article are available in the article.\\

\bibliographystyle{mnras}
\bibliography{bibliography} 

\bsp	
\label{lastpage}







\end{document}